\documentclass[%
 reprint,
 superscriptaddress,
 amsmath,amssymb,
 aps,
 pra,
]{revtex4-1}

\usepackage{graphicx}
\usepackage{dcolumn}
\usepackage{bm}
\usepackage{hyperref}
\usepackage{url}
\usepackage{comment}

\newcommand{\eref}[1]{Eq.~(\ref{#1})}
\newcommand{\Eref}[1]{Equation~(\ref{#1})}
\newcommand{\fref}[1]{Fig.~\ref{#1}}
\newcommand{\Fref}[1]{Figure~\ref{#1}}

\begin{document}
\newcommand{\ManuscriptTitle}{
    Stable Energy Distribution of Weakly Dissipative Gasses\\ under Collisional Energy Cascades
}

\title{\ManuscriptTitle}


\author{Keisuke Fujii}
\email{fujii@me.kyoto-u.ac.jp}
\author{Jun Imano}
\author{Arseniy Kuzmin}
\author{Taiichi Shikama}
\author{Masahiro Hasuo}
\affiliation{%
Department of Mechanical Engineering and Science,
Graduate School of Engineering, Kyoto University
Kyoto 615-8540, Japan}

\date{\today}

\begin{abstract}
Collisional thermalization of a particle ensemble under the energy dissipation can be seen in variety of systems, such as heated granular gasses and particles in plasmas.
Despite its universal existence, analytical descriptions of the steady-state distribution have been missing.
Here, we show that the steady-state energy distribution of the wide class of collisional energy cascades can be well approximated by the generalized Mittag-Leffler distribution, which is one of stable distributions.
This distribution has a power-law tail, as similar to Levy's stable distribution, the index of which is related to the energy dissipation rate.
We demonstrate its universality by comparing Mont-Carlo simulations of dissipative gasses as well as the spectroscopic observation of the atom velocity distribution in a low-temperature plasma.
\end{abstract}

\maketitle

Nonthermal energy distributions have been observed in many nonequilibrium systems.
One of typical classes is the collisional energy cascade, where a high-temperature particle is injected at a certain rate into an ensemble of particles, and this energy is distributed to many other particles through collisions and consumed by the dissipation process in the system.
Because of its simplicity, many systems are categolized into this class.
One example is the granular gasses, which have been modeled with their inelasticity of the collision~\cite{aranson_patterns_2006,villani_mathematics_2006}.
With the energy input, a nontrivial steady state is formed \cite{Rouyer2000,VanZon2004,ben-naim_stationary_2005,Kang2010}, where the input energy is balanced with the energy dissipation by the inelastic collisions.
Another example is an ensemble of atoms in plasmas.
Energetic atoms are generated by several processes~\cite{Corrigan1965,Hey2004,Scarlett2017,McConkey2008,Starikovskiy2015} and this input energy is balanced with collisional energy loss with other particles and walls having lower temperature.

Despite the simplicity and universal existence of the collisional cascade, only a few of its statistical properties have been revealed.
Ben-Naim et al. have pointed out that the velocity distribution of granular gasses under the collisional energy cascade shows a power-law tail~\cite{ben-naim_stationary_2005,Kang2010}.
Although they have reported the analytical representation of the power-law index, the shape of the overall velocity distribution including the low-energy part is not clarified.
Particularly, the convergence to the Maxwell distribution in the no-dissipation limit is not obvious as the power-law index approaches to a finite value in their representation.

The non-thermal velocity distribution of atoms in plasmas have been observed experimentally for a long time~\cite{Vrhovac1991,Amorim2000,Samm1989,Hey1999,Shikama2004}.
As the existence of the high-energy tail significantly changes the chemical reaction rates~\cite{wakelam_kinetic_2012,wakelam_2014_2015,flower_rotational_1997}, the understandings of the energy distribution are demanded~\cite{adamovich_2017_2017}.
Many groups have empirically approximated such non-thermal energy distributions by a sum of a few Maxwell distributions~\cite{Amorim2000, Hey1999}. 
Some Monte-Carlo simulations have reproduces the observed non-thermal velocity distribution~\cite{Sommerer1991,Starikovskiy2015,Ponomarev2017}. 
However, the analytical representation is still missing and thus it is difficult to extract physical knowledge from the observation.

In this Letter, we point out that the generalized Mittag-Leffler (G-ML) distribution well approximates the steady-state energy distribution in these collisional energy cascades.
Although the G-ML distribution has no analytic representation except for a few special cases, its Laplace transform can be simply written by $\mathcal{L}_{f_{\operatorname{G-ML}}}(s) = \int_0^\infty f_{\operatorname{G-ML}}(E)\; e^{-sE} dE = [1 + (\epsilon_0 s)^\alpha]^{-\nu}$. 
Here, $\epsilon_0 > 0$  is the energy scale, $\nu > 0$ is related to the degree of freedom of the system, and $0 < \alpha \leq 1$ is the stability parameter related to relative importance of the dissipation process.
The G-ML distribution has a power-law tail, $f_{\operatorname{G-ML}}(E) \approx \nu \alpha \epsilon_0^\alpha E^{-\alpha-1} / \Gamma(1-\alpha)$, where $\Gamma(x) = \int_0^\infty t^{x-1} e^{-t} dt$ is the Gamma function. 
The existence of the power-law energy tail is consistent with the previous reports~\cite{ben-naim_stationary_2005,Kang2010}, but at the same time this distribution naturally converges to the Maxwell distribution as $\alpha \rightarrow 1$.

\begin{figure*}
    \includegraphics[width=17cm]{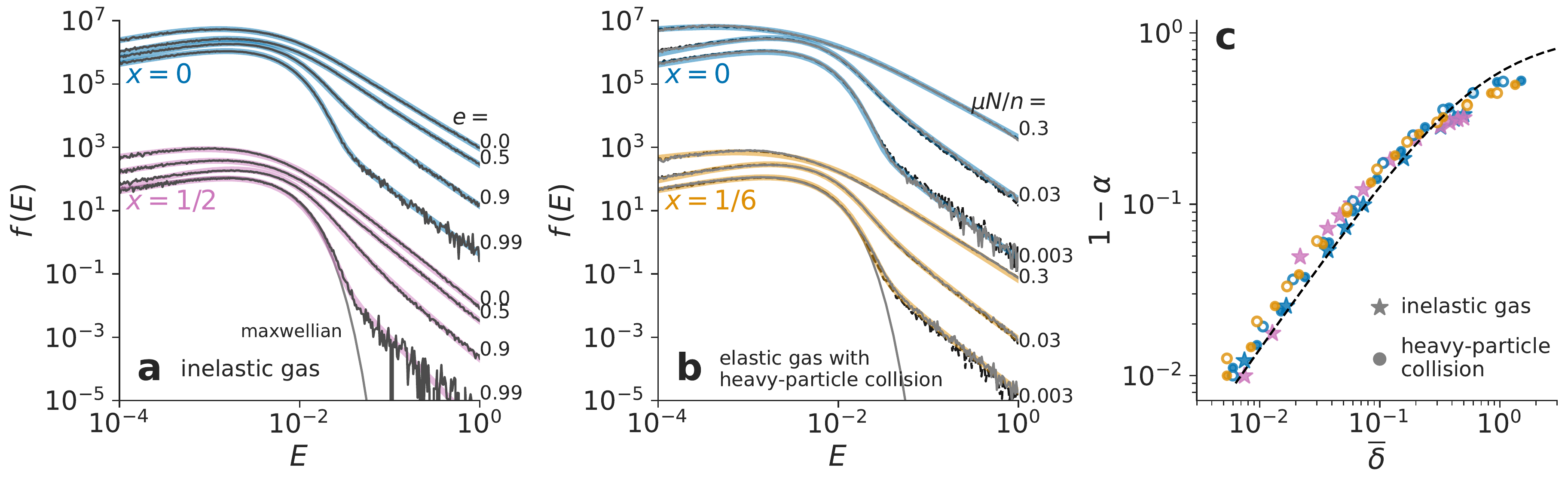}
    \caption{%
    The steady-state energy distribution with the collisional energy cascade computed by the Monte-Carlo simulation.
    (a)~The energy distributions of inelastic gas.
    The upper and lower curves show the results for the Maxwell gas ($x=0$) and the classical gas ($x=1/2$), respectively.
    The restitution coefficients used for the simulations are $e=$ 0.99, 0.9, 0.5, and 0.0.
    The best fits by the G-ML distribution are shown by bold curves. 
    The values of $\alpha$ estimated from the best fit is shown in (c) by stars as a function of $\overline{\delta}$, which is calculated from \eref{eq:delta_inelastic_gas}.
    (b)~The energy distributions of the elastic gas colliding with heavier species.
    The upper and lower curves show the results for the Maxwell gas ($x=0$) and the Van-der-Waals gas ($x=1/6$), respectively.
    Thin solid and dashed curves (they are almost identical and hardly distinguishable) show the results with different mass ratio $\mu = 10^{-1}$ and $10^{-2}$, respectively.
    Bold curves are the best fits by the G-ML distribution.
    The best-fit $\alpha$ is shown in (c) by circles as a function of $\overline{\delta}$, which is calculated from \eref{eq:delta_plasma}.
    The theoretical prediction of $\alpha$ (\eref{eq:alpha_nu}) is show by dashed curves in (c). 
    All the estimated values of $\alpha$ are consistent with the theory.
    }
    \label{fig:simulation}
\end{figure*}

Throughout this Letter, we focus on spatially homogeneous and isotropic systems.
We start from the kinetics of Maxwell-particles in $d$-dimensional space with no energy dissipation.
Let $f(E)$ be the number density of the particle with the kinetic energy $E$. 
As we assume the Maxwell particle, the collision rate among them has no $E$-dependence~\cite{maxwell_iv_1867}. 
Therefore, we can consider the energies of two colliding particles as random samples from $f(E)$.
Let $E_1$ and $E_2$ be the energies of the colliding particles.
After the collision, the total kinetic energy $E_1 + E_2$ are partitioned and distributed to the two particles.
With the partition ratio $u \in [0, 1]$, which is also a random variable, the post-collision energy of one particle $E'$ can be written as
\begin{align}
    \label{eq:recursive}
    E' = (E_1 + E_2) u.
\end{align}
At the steady state, $E'$ also follows $f(E)$, i.e., $f(E)$ is one kind of stable distributions against the operation \eref{eq:recursive}.

The distribution of $u$, $p(u)$, should be symmetric against the exchange of $u$ and $(1-u)$.
From the statistical weight of the $d$-dimensional space $\propto E^{d/2-1}$, we expect the beta distribution $B_{d/2, d/2}(u) = u^{d/2-1} (1-u)^{d/2-1} / B(d/2,d/2)$ as a reasonable choice for the partition distribution $p(u)$.
Here, $B(a, b) = \int_{0}^{1}x^{a-1}(1-x)^{b-1}dx$ is the beta function.
Note that in this form of $p(u)$ we implicitly assume the no-memory limit, where the energies of the two colliding particles are completely randomized by the collision and thus $u$ does not depend on $E_1$ nor $E_2$.
See Supplemental Material for the correction of this assumption\footnote{For the theoretical and experimental details, see the Supplemental Material, which includes \cite{kremer_chapman-enskog_2010,fantz_spectroscopypowerful_2006,wood_extension_1920,McNeill1982,Baravian1987,petrovic_excitation_1992,Cvetanovic2009} as references}. 

Because at this point we have no dissipation processes, $f(E)$ should be a Maxwellian.
This is easily seen by taking the Laplace transform of the distribution, $\mathcal{L}_f(s) = \int_0^\infty f(E)e^{-sE}dE$;
the sum of two random variables (i.e., the convolution) is written as a product of $\mathcal{L}_f$, and the multiplication of the random variables can be written as a scale mixture. 
\Eref{eq:recursive} is equivalent with
\begin{align}
    \label{eq:laplace_simplest}
    \mathcal{L}_f(s) = 
    \int_0^1 \bigl[\mathcal{L}_f(us)\bigr]^2 \;
    p(u)\; du.
\end{align}
Indeed, the Maxwell distribution (the Laplace transform of which is $(1 + \epsilon_0 s)^{-d/2}$) satisfies \eref{eq:laplace_simplest}.

Let us additionally consider an energy dissipation process.
We introduce another random variable $\delta \geq 0$ that represents the fractional energy dissipation per one self-collision.
The energy of our particle after one self-collision and the dissipation process is $E' = (E_1 + E_2) u'$ with $u' = u e^{-\delta}$.
Because of the dissipation process, the distribution $p(u')$ is not symmetric anymore and is skewed toward $u'=0$.
We find that for many dissipation processes this skewed distribution can be approximated by the generalized beta distribution~\cite{Note1},
\begin{align}
    \label{eq:genbeta}
    p(u') = \frac{\alpha}{B(\nu, \nu)}u'^{\alpha \nu - 1} (1-u'^\alpha)^{\nu - 1},
\end{align}
where $0 < \alpha \leq 1$ represents the asymmetricity of the distribution. 

The distribution stable against the operation $E'=(E_1 + E_2) u'$ with \eref{eq:genbeta} is the generalized Mittag-Leffler (G-ML) distribution, the Laplace transform of which is
\begin{align}
    \label{eq:gml}
    \mathcal{L}_{f_{\operatorname{G-ML}}}(s) = \left[1 + (\epsilon_0 s)^\alpha\right]^{-\nu}.
\end{align}
Although the G-ML distribution has no analytic representation, the precise numerical approximation is available~\cite{haubold_mittag-leffler_2011,barabesi_new_2016,korolev_mixture_2020, Note1}.
The G-ML distribution has a power-law tail at large $E$, $f_{\operatorname{G-ML}}(E) \approx \nu \alpha \epsilon_0^\alpha E^{-\alpha-1} / \Gamma(1-\alpha)$, which is consistent with the observation by Ben-Neim et al.~\cite{ben-naim_stationary_2005,Kang2010}.
With the no-dissipation limit ($\alpha \rightarrow 1$ and $\nu \rightarrow d/2$), the G-ML distribution naturally converges to the Maxwell distribution.

The values of $\alpha$ and $\nu$ may depend on dissipation processes.
However, we find that, in many dissipative systems with $\delta \ll 1$, they are represented by the following forms~\cite{Note1}
\begin{align}
    \label{eq:alpha_nu}
    \alpha = \frac{1}{1 + \overline{\delta} / \log 2},
\end{align}
and 
$\nu = \frac{d/2}{\alpha^2 + (1 - \alpha)^2 d/2}$,
where $\overline{\delta}$ is the mean of the fractional energy dissipation.

These discussions can be easily extended to more general gasses, where the collision rate has the energy dependence of $\propto E^{x}$.
For example, the collision rate of the classical balls is proportional to $E^{1/2}$; the particles interacting through the Van-der-Waals potential collide with $E^{1/6}$~\cite{Massey1934, Flannery2006}.
In order to deal with such systems, we consider the weighted distribution, $\hat{f}(E) = E^x f(E) / Z$, with $Z$ the normalization constant.
Based on an approximation $(E_1 / \epsilon_0 + E_2 / \epsilon_0 )^x \approx (E_1 E_2 / \epsilon_0^2)^x$, which is valid if $|x| \ll d/2$, this weighting approximately represents the energy dependence of the collision rate.
Although this weighting changes the statistical weight of the $d$-dimensional space to $\propto E^{d/2 + x - 1}$,
the rest of the discussion is the same and we find
\begin{align}
    \nu = \frac{d/2 + x}{\alpha^2 + (1-\alpha)^2 (d/2 + x)}.
    \label{eq:nu}
\end{align}

As a numerical demonstration of the above discussion, we carried out several Monte-Carlo simulations, one of which is similar to that presented by Ben-Neim et al~\cite{ben-naim_stationary_2005}.
We prepared $10^4$ inelastic particles with the restitution coefficient $e \in [0, 1]$.
With a certain rate we heat one of the particles to the temperature $1$, while this input energy is balanced with the energy dissipation by the inelastic collisions.
A pair of colliding particles are chosen based on their velocities. 
Their post-collision velocities are determined by the differential cross section and the restitution coefficient~\cite{villani_mathematics_2006, Note1}.
Note that in this simulation, we do not assume the non-memory limit of the collision, but we consider the actual collision process taking the momentum conservation and the collision geometry into account.
We ran two simulations, one is for the Maxwell gas ($x=0$) and the other is for the classical gas ($x=1/2$).

In \fref{fig:simulation} (a), we show the steady-state energy distributions of these inelastic particles.
The distribution with $e=0.99$, which is close to the elastic limit, has a similar profile to the Maxwellian in the low-energy region but has a power-law tail in the high-energy side.
With more inelasticity (smaller $e$ value), the power-law tail becomes larger and the low-energy distribution deviates from the Maxwellian more significantly.
The bold curves in the figure show the best fits by the G-ML distributions with taking \eref{eq:nu} into account but with $\alpha$ being left adjustable.
All the results in both the low- to high-energy sides are well represented by the G-ML distribution.

For the hard-sphere inelastic gasses in 3-dimensional space, we find~\cite{Note1}
\begin{align}
    \overline{\delta} \approx \frac{1-e^2}{8(1 + e/3)}.
    \label{eq:delta_inelastic_gas}
\end{align}
Star markers in \fref{fig:simulation} (c) shows the value of $\alpha$ that gives the best fit to the simulated distribution, as a function of $\overline{\delta}$ which is computed from the value of $e$ used in the simulation. 
The values of $\alpha$ at the best fit are consistent with \eref{eq:alpha_nu} (the dashed curve).

Another simulation was carried out to study the atom kinetics in plasmas, where the atoms experience elastic collisions not only among the same species but also with heavier particles [we can also consider collision to a material surface instead of the heavy particles].
We assume that the temperature of the heavier particles is zero and unchanged, i.e., the heavy-particle collision acts as the energy dissipation.
With this setting, the mean of the fractional energy dissipation $\delta$ can be written as~\cite{Note1}
\begin{align}
    \overline{\delta} = 2^{1-x}\mu\frac{N \sigma^N_\mathrm{mt}}{n \sigma^n_\mathrm{vi}},
    \label{eq:delta_plasma}
\end{align}
where $\mu = mM/(m+M)^2 \ll 1$ is the reduced mass ratio ($m$ and $M$ are the masses of our atoms and heavier particles, respectively), 
$n$ is the density of our atoms, $N$ is the density of the heavy particles, and $\sigma^n_\mathrm{vi}$ and $\sigma^N_\mathrm{mt}$ are the viscosity cross section of the self-collision and the momentum transfer cross section of the heavy-particle collision, respectively.

Similar to the previous simulation, we prepare $10^4$ atoms, generate a high energy atom with the temperature $1$ at a certain rate, and simulate their collisional energy cascade.
We assume the isotropic-type collision~\cite{Note1} and use the identical cross section for the self-collision and the heavy-particle collision.
Two simulations with the Maxwell gas ($x=0$) and the Van-der-Waals gas ($x=1/6$) were carried out.

\Fref{fig:simulation} (b) shows the steady-state energy distributions of these atoms in plasmas.
The results with several values of $\mu N /n$ (0.003, 0.03 and 0.3) are shown in the figure.
Also, the simulated distributions under the two $\mu$ values ($10^{-1}$ and $10^{-2}$) but with the same $\mu N /n$ are plotted by black and gray curves.
They are almost identical and hardly distinguished, as predicted by \eref{eq:delta_plasma}.
As similar to the inelastic-gas simulation, all the distributions have a power-law tail.
With more dissipation, the tail becomes more significant.

The bold curves in the figure show the best fit by the G-ML distributions with taking \eref{eq:nu} into account.
The G-ML distribution well reproduces the results.
The circle markers in \fref{fig:simulation}~(c) show the values of $\alpha$ at the best fit as a function of $\overline{\delta}$, which is computed from \eref{eq:delta_plasma} and the value of $\mu N/n$ used in the simulation. 
Results with $\mu = 10^{-1}$ and $10^{-2}$ are shown by filled and open circles, respectively.
All of them are close to the theoretical prediction, \eref{eq:alpha_nu}.

\begin{figure}
    \centering
    \includegraphics[width=8cm]{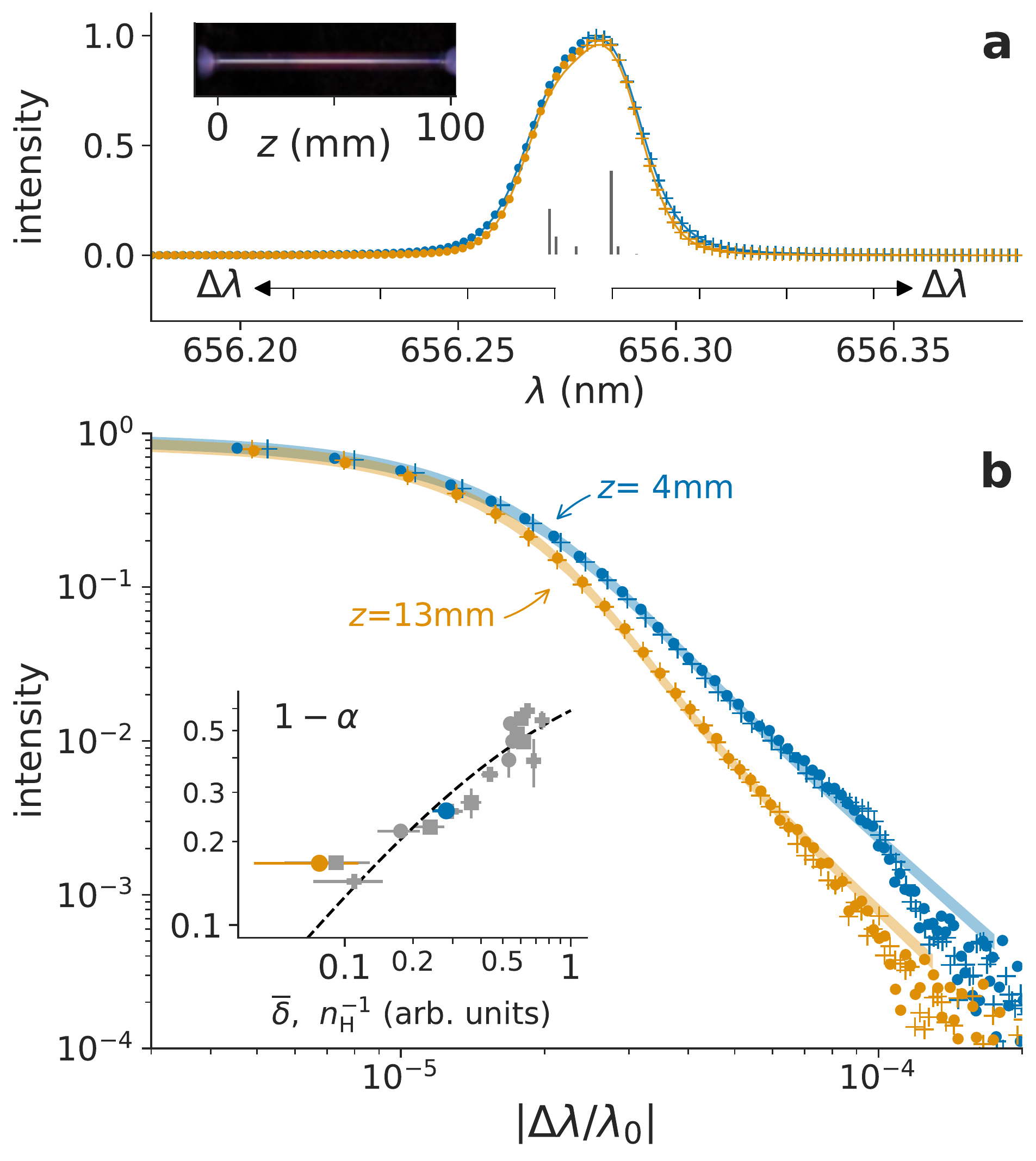}
    \caption{%
        Hydrogen Balmer-$\alpha$ spectra observed from the low-pressure glow discharge tube at different positions.
        The inset of (a) shows the discharge tube and the $z$-axis along the capillary.
        (b) The spectra in the log-log plot as a function of the wavelength shift $\Delta\lambda$, where $\Delta\lambda$ is defined as the wavelength shift from the nearest fine structure (see the axis in (a)).
        The solid curves are the Gaussian-convoluted G-ML distribution taking the fine structure into account.
        In the inset of (b), the $n_\mathrm{H}$-dependence of the best-fit $\alpha$ values are shown, where $n_\mathrm{H}$ is estimated from the intensity ratio of the atomic and molecular lines of hydrogen~\cite{Note1}.
    }
    \label{fig:spectra}
\end{figure}

As an experimental demonstration, we carried out spectroscopic observations of a low-temperature hydrogen plasma generated inside a commercially available discharge lamp (Edmund optics, \#60-906, shown in the inset of \fref{fig:spectra}~(a)) with a DC high-voltage power supply (Spellman SP16122A)~\cite{Note1}.
Hydrogen atoms in the discharge plasma are mainly generated through electron-impact dissociation, where the generated atoms typically have 3 eV kinetic energy~\cite{Corrigan1965,Hey2004,Scarlett2017}.
This kinetic energy is distributed to other atoms by self-collisions and dissipated by the collision with the walls and other species.
We observed the hydrogen atom Balmer-$\alpha$ line (central wavelength $656.280$ nm) with a homemade high-resolution spectrometer with the wavelength resolution of 8 pm at $\lambda \sim 650$ nm.
The observation was carried at different positions $z$ of the lamp capillary (see \fref{fig:spectra}~(a) inset).
The small inner diameter of the capillary (1 mm) compared with the mean free path of atoms makes the diffusion coefficient small.
Therefore, the dissociation ratio is higher at deeper position of the capilarly.
This tendency was confirmed by the intensity ratio against the molecular Fulcher-$\alpha$ line~\cite{Note1}.

\Fref{fig:spectra}~(a) shows the observed Balmer-$\alpha$ spectrum at $z = 4$ and 13 mm.
In \fref{fig:spectra}~(b), we show the same spectrum in the log-log plot as a function of the wavelength shift.
Note that since the Balmer-$\alpha$ mainly consists of two components, we define the wavelength shift $\Delta \lambda$ from the nearest line center as shown in \fref{fig:spectra}~(a).
The wing profile is far from the Gaussian but rather having power-law tails.
A steeper tail (i.e., closer to the Gaussian) is found in the spectrum measured at $z$ = 13 mm than that at $z$ = 4 mm, while the dissociation ratio at $z = $ 13 mm ($\approx 10^0$) is larger than that in $z = 4$ mm ($\approx 10^{-1}$).
This spectral profile mostly reflects the velocity distribution of hydrogen atoms through the Doppler effect.
We note that although the Lorentz component due to the Stark effect and the diffraction effect inside the spectrometer~\cite{Fujii2014a}, both of which scale $|\Delta \lambda|^{-2}$ in the tail, are small, we estimated them from the further wing region of the observed spectra and subtracted.

The interaction between two hydrogen atoms can be approximated by the Van-der-Waals potential, thus $x=1/6$~\cite{international1999iaea,chapman_mathematical_1991,Note1}.
A hydrogen atom also collides with the inner wall of the capillary and hydrogen molecules, which should act as the energy dissipation.
In our experiment they should have a finite temperature, though we have assumed the zero temperature for the heavier species in the above theory.
The instrumental broadening (which is close to the Gaussian profile) also affects the observed profiles.
Therefore, we consider the convolution of the Gaussian distribution and the velocity distribution derived from the G-ML energy distribution~\cite{Note1}.

The best fit with the Gaussian-convoluted G-ML distribution is shown by solid curves in \fref{fig:spectra}~(b).
The observed profiles are well represented by this distribution up to $|\Delta \lambda / \lambda_0 | \lesssim 7 \times 10^{-4}$, which corresponds to 9 eV kinetic energy.
The values of $\alpha$ at $z$ = 13 mm and 4 mm are 0.83 and 0.74, respectively.

In the inset of \fref{fig:spectra}~(b), we show the estimated values of $\alpha$ as a function of the atom density $n_\mathrm{H}$, which is estimated from the dissociation ratio~\cite{Note1}.
Here, we assume that the $\overline{\delta}$ is inversely proportional to $n_\mathrm{H}$.
Furthermore, since neither the absolute atom density nor the restitution coefficient of the wall are clear, we scaled the horizontal values with a single constant to match the theoretical prediction~\eref{eq:alpha_nu}.
The positive dependence of $1 - \alpha$ on $\overline{\delta}$ is consistent with the theoretical prediction.

\begin{figure}
    \centering
    \includegraphics[width=8cm]{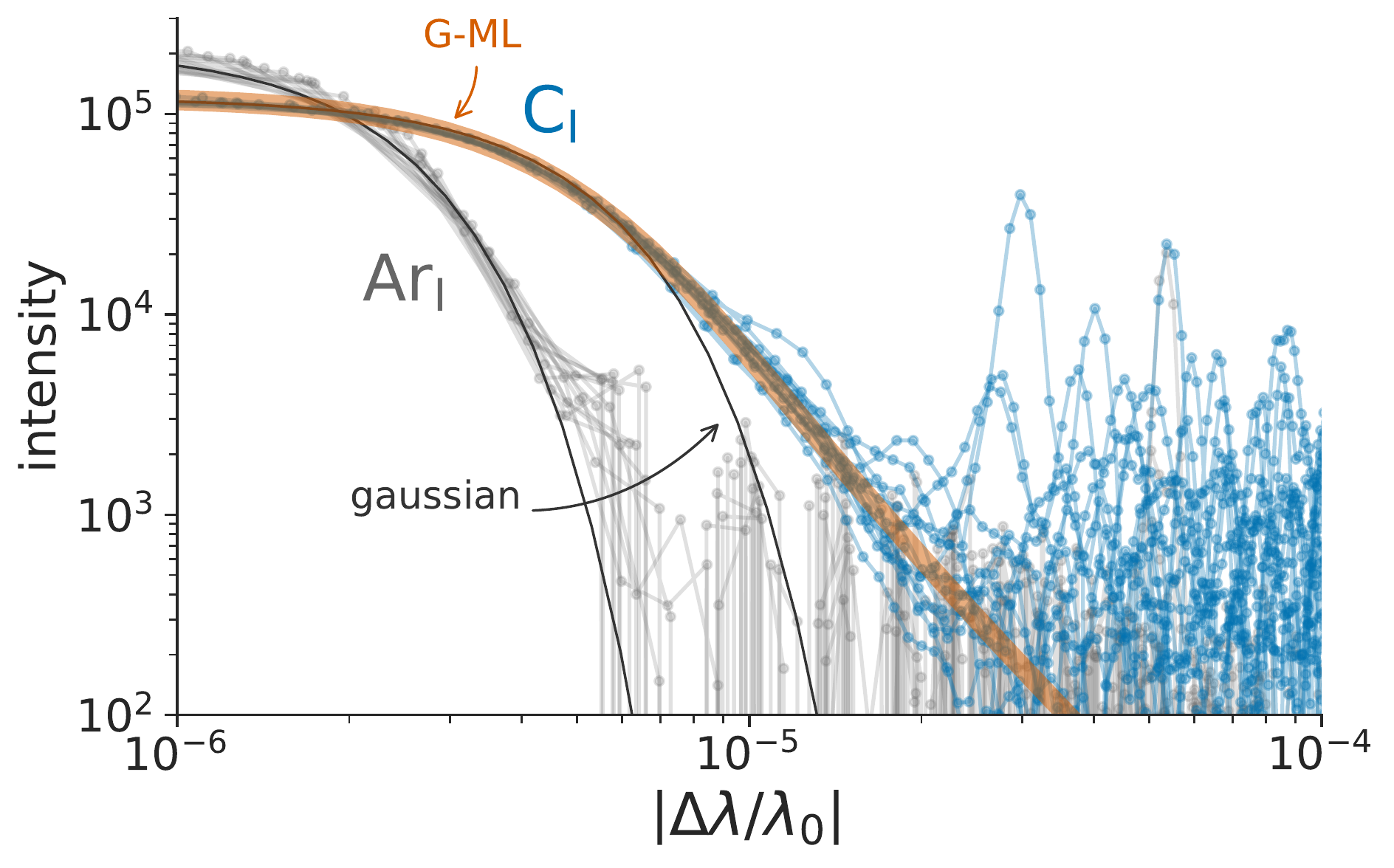}
    \caption{%
        Emission spectra of neutral carbon atoms measured for carbon-nitrogen-argon plasma at the National Solar Observatory~\cite{NSO}.
        $\lambda_0 = $ 909.483, 1068.308, 1068.535, 1069.125, 1070.734, 1072.953 and 1454.249 nm lines are plotted with the area normalized to unity.
        The thin and bold curves show the best fit by the Gaussian and the Gaussian-convoluted G-ML distribution, respectively.
        Eight neutral argon emission spectra 
        are also shown by gray points, which indicate the negligible effect of the instrumental profile on the carbon profiles.
    }
    \label{fig:other_spectra}
\end{figure}

This G-ML velocity distribution is also expected for other atoms.
\Fref{fig:other_spectra} (b) shows the neutral carbon atom emission profiles measured at the Kitt Peak Solar Observatory~\cite{NSO} from a microwave discharge generated with the mixture of carbon, nitrogen, and argon to study cyanide spectra.
The spectrum in $\lambda=$ 830--2710 nm was observed with a high-resolution Fourier transform spectrometer at the observatory.
In the figure, we show seven different emission lines of neutral carbon measured simultaneously as a function of $|\Delta \lambda /\lambda_0|$.
All the profiles show a similar distribution, indicating the dominant contribution of the Doppler broadening over other broadening effects, such as Stark broadening or pressure broadening.
From the neutral argon emission lines observed at the same time (gray points in \fref{fig:other_spectra}), we find that the instrumental broadening has a negligible effect on the carbon line profiles.
As seen in the figure, the carbon profiles show significant tails compared with a Gaussian function (thin curve in \fref{fig:other_spectra}).
From the fit by the Gaussian-convoluted G-ML distribution (bold curve in the figure), we obtain $\alpha = 0.77 \pm 0.02$, suggesting significant self-collisions in this plasma.

In this Letter, we show that the G-ML distribution universally represents the steady-state energy distribution of the collisional energy cascade.
It is validated by the numerical simulation for the dissipative gasses, as well as the spectroscopic observation of the velocity distribution of atoms in plasmas.
Because the stationary parameter $\alpha$ depends on the energy dissipation ratio, i.e., how close the system is to the thermal equilibrium, this provides a new diagnostic tool for various systems.

Here, we discussed a fundamental property of spatially homogeneous and isotropic systems, and thus further investigations are necessary to apply our findings to non-uniform systems~\cite{Note1}.
However, our assumptions may be valid if the mean free path of the particles is comparable to or larger than the system size where the spatial temperature gradient is not significant, such as the divertor plasmas of nuclear fusion reactors.
Although typical divertor simulation codes assume the Maxwellian for the local velocity distributions of atoms to compute their transport~\cite{kukushkin_effect_2005}, the G-ML distribution can be another reasonable choice for such systems.

\begin{acknowledgments}
    This work was supported by JSPS KAKENHI Grant Number 19K14680 and 19KK0073.
    K. F. thanks Dr. Keiichiro Urabe for the fruitful discussions.
    K. F. also thanks an anonimous person with the username \textit{vitamin d}, who gave us an essential suggestion in \texttt{mathoverflow} https://mathoverflow.net/questions/401835/inverse-laplace-transform-of-frac1sa-1-with-0-a-leq-1
\end{acknowledgments}

\bibliography{refs}

\pagebreak

\newcommand{\ManuscriptTitleSupp}{\ManuscriptTitle}
\newcommand{\EqRecursive}{\ref{eq:recursive}}
\newcommand{\EqGenBeta}{\ref{eq:genbeta}}
\newcommand{\EqAlpha}{\ref{eq:alpha_nu}}
\newcommand{\EqNu}{\ref{eq:nu}}
\newcommand{\EqDeltaInelastic}{\ref{eq:delta_inelastic_gas}}
\newcommand{\EqDeltaPlasma}{\ref{eq:delta_plasma}}

\widetext
\begin{center}
\textbf{\large \ManuscriptTitleSupp}
\end{center}
\setcounter{equation}{0}
\setcounter{figure}{0}
\setcounter{table}{0}
\makeatletter

\renewcommand{\theequation}{S\arabic{equation}}
\renewcommand{\thefigure}{S\arabic{figure}}
\renewcommand{\thetable}{S\arabic{table}}

\section{Theoretical Details}

\subsection{Elastic Gas with Heavy-Particle Collision}
In this subsection, we consider the elastic gas undergoing self-collisions and heavy-particle collisions.
The densities of our particle and the heavy particle are $n$ and $N$, respectively.
In the following, we consider the kinetic energy loss and randomization based on the collision theory.
Cross sections for the self-collision and the heavy-particle collision are distinguished by the superscripts $n$ and $N$, respectively.

\begin{figure*}[b]
    \centering
    \includegraphics[width=10cm]{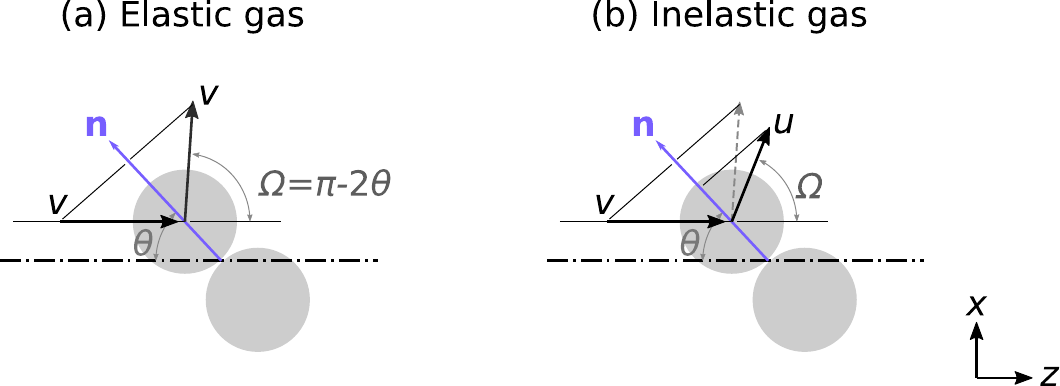}
    \caption{
        Schematic illustration of the two-body collision in the center-of-mass frame.
        (a) Elastic collision and (b)inelastic collision.
    }
    \label{sup:fig:collision}
\end{figure*}

\subsubsection{Energy loss by heavy-particle collision}

Let us consider an atom with mass $m$ moving toward the $z$ axis with velocity $\mathbf{u} = (0, 0, u_z)$ and elastically colliding with a heavier particle at rest with mass $M \gg m$.
The center of mass (CM) velocity is written as 
$\mathbf{V}_\mathrm{CM} = \frac{m}{m+M} \mathbf{u}$.
With the scattering angle $\Omega$ in the CM frame, the atom velocity after the elastic collision can be written as follows:
\begin{align}
    v_x &= (u - V_\mathrm{CM}) \sin \Omega\\
    v_z &= V_\mathrm{CM} + (u - V_\mathrm{CM}) \cos \Omega.
\end{align}
Here, we assume that the scattering takes place in the $x$-$z$ plane without any loss of generality.
The amount of the kinetic energy of our particle lost by the collision is
\begin{align}
    \label{sup:eq:scattering}
    \epsilon &= \frac{m}{2}(v_x^2 + v_z^2) = 
    \frac{m}{2} u^2 \left(1 - \frac{2mM}{(m+M)^2}(1-\cos \Omega)\right) = 
    E\left(1-2\mu(1-\cos \Omega)\right),
\end{align}
with $\mu = \frac{mM}{(m+M)^2}$.
By integrating over the scattering angle $\Omega$, we obtain the average value of the energy loss.
In the 3-dimensional space, this is
\begin{align}
    1-\frac{\overline{\epsilon}}{E}
    = 2\mu\int_0^\pi (1 - \cos \Omega) \frac{\partial\sigma^N(\Omega)}{\partial \Omega} 2\pi \sin\Omega d\Omega 
    = 2\mu \sigma^N_\mathrm{mt},
\end{align}
where, $\partial\sigma^N(\Omega) / \partial \Omega$ is the differential cross section for this heavy-particle collision. 
Here we see that the mean energy loss can be written with the momentum transfer cross section $\sigma^N_\mathrm{mt}$.
The list of $\sigma_\mathrm{mt}$ for several types of the collisions can be found in Table~\ref{sup:tab:crosssections}.

\begin{figure*}
    \centering
    \includegraphics[width=12cm]{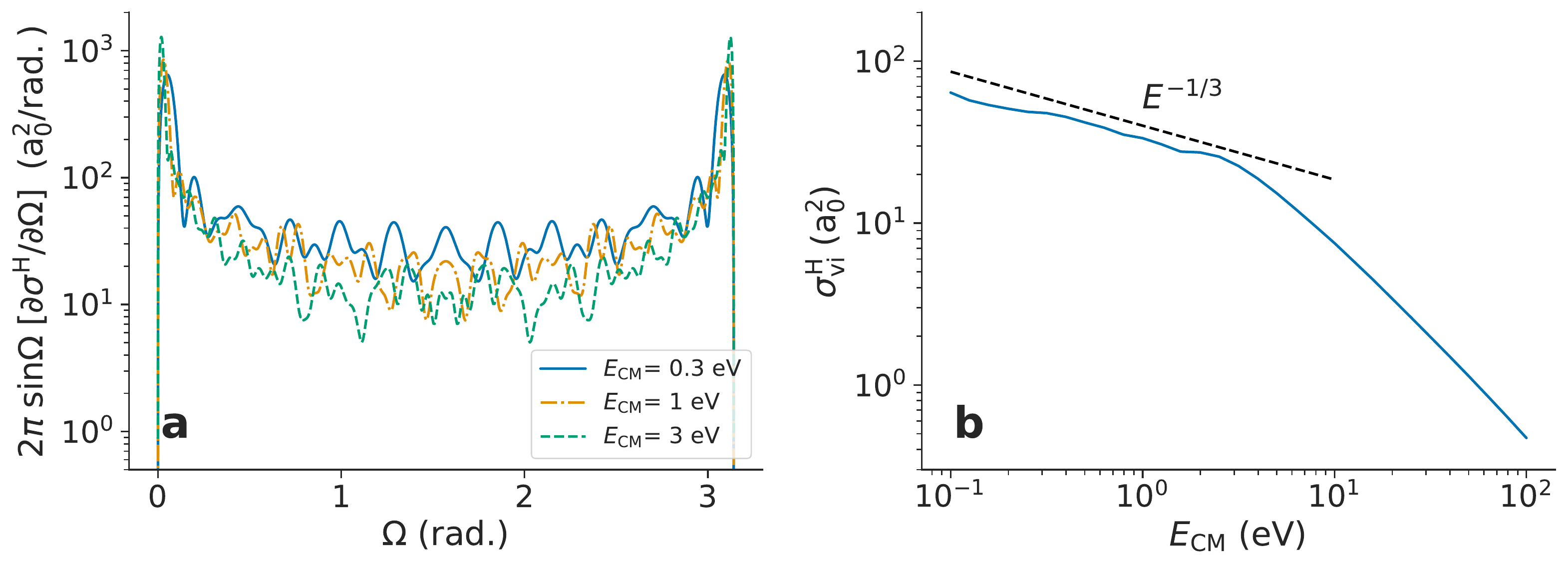}
    \caption{
        Cross section the elastic collision between two hydrogen atoms.
        (a) Differential cross-section and (b) the energy dependence of the viscosity cross section.
    }
    \label{sup:fig:crosssection}
\end{figure*}

\subsubsection{The energy randomization}
In Eq.~(\EqRecursive), we assume that in one self-collision the kinetic energies of two collided particles are completely randomized with keeping the total kinetic energy $E_1 + E_2$.
However, the randomization of the kinetic energies is not generally perfect.
For example, the collision with $\Omega \approx \pi$ does not change their velocity significantly and with $\Omega \approx 0$ only swaps their energies.
As an example of the collision of quantum particles, we show the differential cross section of elastic collision between two hydrogen atoms in \fref{sup:fig:crosssection}~(a).
The significance of the forward ($\Omega\approx \pi$) and backward ($\Omega \approx 0$) parts of the cross section is aparent.

In order to quantify the randomization efficiency for each collision, the viscosity cross section $\sigma_\mathrm{vi}$ has been used~\cite{kremer_chapman-enskog_2010}  
\begin{align}
    \sigma^n_\mathrm{vi}
    = \int_0^\pi \sin^2\Omega\;\; \frac{\partial\sigma^n(\Omega)}{\partial \Omega} 2\pi \sin\Omega d\Omega,
\end{align}
where 
$\frac{\partial\sigma^n(\Omega)}{\partial \Omega}$ is the differential cross section for the self-collision.
Table~\ref{sup:tab:crosssections} shows the value of $\sigma_\mathrm{vi}$ for several types of the collisions.
\Fref{sup:fig:crosssection}~(b) shows the viscosity cros ssection of hydrogen atom-atom elastic collision.
As the Van-der-Waals interaction is dominant at $E \lesssim 3$ eV, the cross section behaves as $\propto E^{-1/3}$.
Note that in general, the cross section of the elastic collision of two particles interacting with $r^{-a}$-potential has the energy dependence of $\propto E^{-a/2}$~\cite{chapman_mathematical_1991}, where $r$ is the internuclear distance.

\subsubsection{Average of the fractional energy dissipation $\overline{\delta}$}

The mean of the fractional energy dissipation $\overline{\delta}$ for the elastic gas with heavy particle collision is
\begin{align}
    \overline{\delta} = \left(1-\frac{\overline{\epsilon}}{E}\right) \frac{N}{n\sigma^n_\mathrm{vi}}
    = 2\mu 
    \frac{N \sigma^N_\mathrm{mt}}{2^x n\sigma^n_\mathrm{vi}}.
    \label{sup:eq:delta_plasma}
\end{align}
Here, we assume that $\sigma^N_\mathrm{mt}$ and $\sigma^n_\mathrm{vi}$ have the same energy dependence, $\propto E^{x-1/2}$.
The term $2^x$ in the denominator comes from the difference in the relative velocity of the self-collision ($\propto (2E)^{1/2}$) and the heavy-particle collision ($\propto E^{1/2}$).

\bgroup
\def\arraystretch{1.5}
\begin{table}
    \caption{\label{sup:tab:crosssections}A list of crosssections for basic elastic collisions in 3-dimensional space.}
    \begin{ruledtabular}
    \begin{tabular}{cccc}
    \textbf{type} & \textbf{differential crosssection} $\frac{\partial \sigma}{\partial \Omega}$ & 
    \textbf{momentum transfer crosssection} $\sigma_\mathrm{mt}$ & 
    \textbf{viscosity crosssection} $\sigma_\mathrm{vi}$\\
    \hline
    hard sphere & $\frac{\sigma_\mathrm{el}}{4 \pi}$ & $\sigma_\mathrm{el}$ & $\frac{2}{3}\sigma_\mathrm{el}$\\ 
    isotropic & $\frac{\sigma_\mathrm{el}}{\pi \sin\Omega}$ & $\sigma_\mathrm{el}$ & $\frac{1}{2}\sigma_\mathrm{el}$
    \end{tabular}
    \end{ruledtabular}
\end{table}
\egroup

\subsection{Inelastic gas}

When two particles undergo an inelastic collision, the scattering angle $\Omega$ depends on the restitution coefficient $0 \leq e \leq 1$ (\fref{sup:fig:collision}~(b)).
Because of the inelasticity, the momentum normal to the collision direction (vector $\mathbf{n}$ in the figure) changes~\cite{villani_mathematics_2006},
\begin{align}
    v \cos \theta = e u \cos \theta,
    \label{sup:eq:pdelta}
\end{align}
with the post-collision velocity $u$ in the CM frame.
The momentum perpendicular to $\mathbf{n}$ is conserved.
$u$ and the scattering angle $\Omega$ can be written as
\begin{align}
    \label{sup:eq:velocity_loss}
    u &= v \sqrt{\sin^2 \theta + e^2 \cos^2 \theta},\\
    \Omega &= \cos^{-1}\left(-\frac{e\cos\theta}{\sqrt{\sin^2 \theta + e^2\cos^2 \theta}}\right) - \theta.
\end{align}

In this work, we use the hard-sphere cross section for the inelastic gas.
By averaging \eref{sup:eq:velocity_loss}, we obtain $ \langle u \rangle = v \frac{1-e^2}{2}$.
Since in an isotropic system, the kinetic energy should be shared equally by the kinetic energy in the CM frame and that of the center of mass, i.e., $\langle v^2 \rangle = \langle V_\mathrm{CM}^2 \rangle$,
the average of the fractional energy loss per one collision is $(1-e^2)/4$.

As similar to the discussion for the elastic gas, we use the viscosity cross section to find the effective collision rate.
With the elastic limit $e \rightarrow 1$, we have $\sigma_\mathrm{vi} = 2\sigma_\mathrm{el} / 3$ as shown in Table~\ref{sup:tab:crosssections}.
Here $\sigma_\mathrm{el}$ is the total crosssection.
In the complete inelastic limit $e = 0$,$\sigma_\mathrm{vi} = \sigma_\mathrm{el} / 2$.
Although no analytic forms are available for an arbitrary value of $e$, we adopt the linear approximation $\sigma_\mathrm{vi} \approx 1/2 + e / 6$, which gives Eq.~(\EqDeltaInelastic).

\subsection{Derivation of the generalized beta distribution}

\begin{figure*}[b]
    \centering
    \includegraphics[width=13cm]{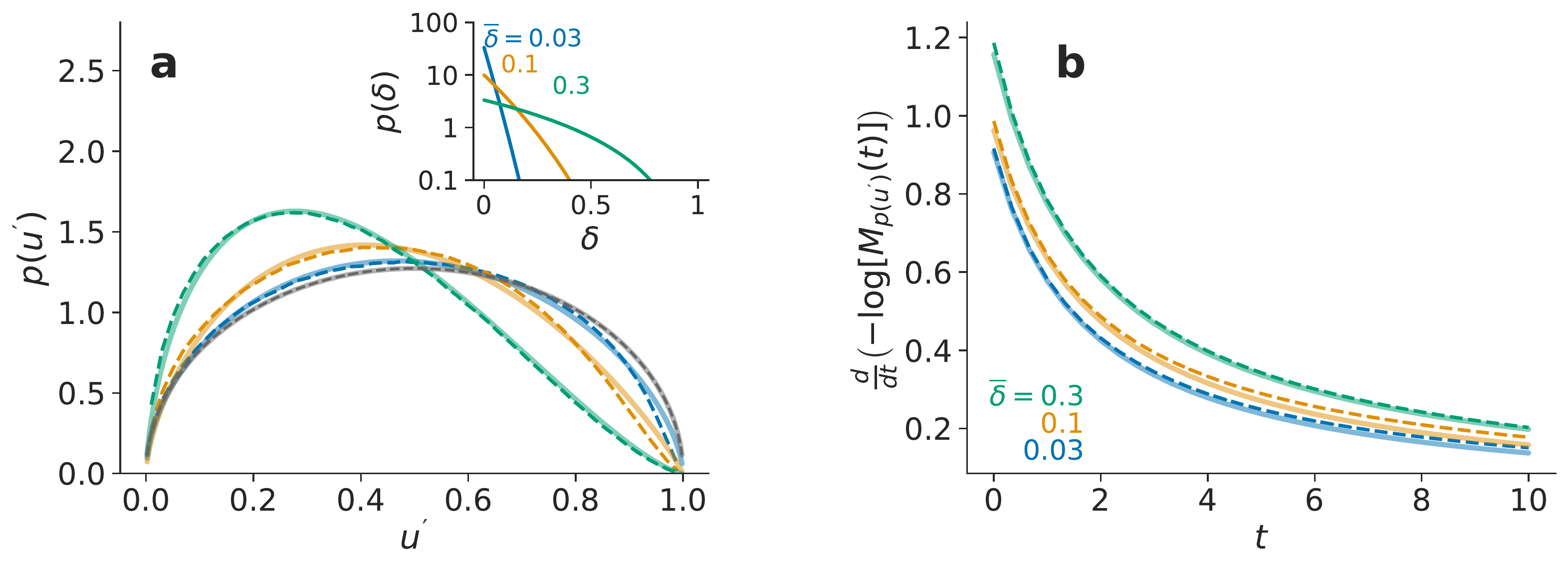}
    \caption{
        Comparison between \eref{sup:eq:u_prime} and Eq.~(\EqGenBeta).
        (a) The distributions of \eref{sup:eq:u_prime} and Eq.~(\EqGenBeta), with $\overline{\delta} = 0.03, 0.1$, and $0.3$ and $d = 3$.
        The dashed curves represent \eref{sup:eq:u_prime} while the solid curves represent Eq.~(\EqGenBeta) with $\alpha$ and $\nu$ determined from Eq.~(\EqAlpha) and Eq.~(\EqNu), respectively.
        The gray curve is the distribution for $\overline{\delta}=0$, i.e., the standard beta distribution, which is symmetric against $u'=1/2$.
        In the inset of (a), the distribution of $\delta$ (\eref{sup:eq:pdelta}) is plotted.
        (b) Comparison of their moments. The values of \eref{sup:eq:gen_beta_moment_deriv} are plotted for $\overline{\delta} = 0.03, 0.1$ and 0.3.
        The dashed curves are from \eref{sup:eq:u_prime} while the solid curves are from Eq.~(\EqGenBeta).        
    }
    \label{sup:fig:genbeta}
\end{figure*}

Let us consider the Maxwell atoms experiencing heavy-particle collisions with $\mu \ll 1$.
The time interval $\tau$ between the two successive self-collision follows $p(\tau) =  e^{-\tau / (n\sigma_\mathrm{vi})} / n\sigma_\mathrm{vi}$.
After one heavy-particle collision, the kinetic energy of atoms changes as a factor of $\approx e^{-2\mu \sigma_\mathrm{mt}^N}$.
Therefore, the distribution of the fractional energy loss $\delta\in[0,1]$ between two successive self-collisions can be written as
\begin{align}
    p(\delta) = \frac{1}{\overline{\delta}}(1-\delta)^{1/\overline{\delta} - 1},
\end{align}
with $\overline{\delta}$ is defined by \eref{sup:eq:delta_plasma}.
The distribution of $u' = u (1-\delta)$ can be derived from the product distribution,
\begin{align}
    p(u') = \frac{\frac{1}{\overline{\delta}} u'^{1/\overline{\delta}-1}}{B\left(\frac{d}{2},\frac{d}{2}\right)}\left[
        B\left(\frac{d}{2}-\frac{1}{\overline{\delta}}, \frac{d}{2}\right) - 
        B\left(u'; \frac{d}{2}-\frac{1}{\overline{\delta}}, \frac{d}{2}\right)
    \right],
    \label{sup:eq:u_prime}
\end{align}
where $B(x; a, b) = \int_x^1 t^{a-1} (1-t)^{b-1} dt$ is the incomplete beta function.
Dashed curves in \fref{sup:fig:genbeta}~(a) show the distribution by \eref{sup:eq:u_prime} for several values of $\overline{\delta}$.
With a larger value of $\overline{\delta}$, the distribution is skewed more significantly.

The solid curves in \fref{sup:fig:genbeta}~(a) show the generalized beta distribution Eq.~(\EqGenBeta).
Both distributions are very close to each other.
In order to find the best values of $\alpha$ and $\nu$ that give the closest profile to \eref{sup:eq:u_prime} with given $\overline{\delta}$, we compare the $t$-th moments around $u'=0$ for both the distributions,
\begin{align}
    \mathcal{M}_{p(u')}(t) =
    \frac{1}{\overline{\delta}t + 1} B\left(\frac{d}{2} + t, \frac{d}{2}\right) \approx 
    B\left(\frac{\nu + t}{\alpha}, \frac{\nu}{\alpha}\right).
    \label{sup:eq:gen_beta_moment}
\end{align}
We take the logarithm of \eref{sup:eq:gen_beta_moment} and differentiate by $t$,
\begin{align}
    \frac{d}{dt}\left[-\log\mathcal{M}_{p(u')}(t)\right] = 
    \frac{1}{t + 1 / \overline{\delta}} + \psi\left(d + t\right) - \psi\left(\frac{d}{2} +t \right)
    \approx 
    \frac{1}{\alpha} \psi\left(2\nu + \frac{t}{\alpha}\right) - \frac{1}{\alpha}\psi\left(\nu + \frac{t}{\alpha}\right),
    \label{sup:eq:gen_beta_moment_deriv}
\end{align}
where $\psi(x) = \frac{d}{dx} [\log\Gamma(x)]$ is the digamma function.
By matching the zero-th and first derivatives of \eref{sup:eq:gen_beta_moment_deriv} at $t=0$ with substituting the approximtion $\psi(x) \approx \log(x) - \frac{1}{2x}$, we obtain Eq.~(\EqAlpha) and Eq.~(\EqNu).
The values of the both sides of \eref{sup:eq:gen_beta_moment_deriv} are shown in \fref{sup:fig:genbeta}~(b) by dashed and solid curves, respectively.
The generalized beta distribution very well approximates \eref{sup:eq:u_prime}.

\subsection{Validity Conditions}

In the main text, we adopted the following assumptions to develop our theory,
\begin{enumerate}
    \item The systems are in the steady state.
    \label{sup:enum:steady}
    \item The systems are spatially homogeneous and isotropic.
    \label{sup:enum:homo}
    \item The collision rate is proportional to $E^x$.
    \label{sup:enum:rate}
    \item The particle source is located at the high-energy side.
    \label{sup:enum:process}
    \item (for the heavy-particle collision system) The heavy-particle temperature is zero.
    \label{sup:enum:zero}
    \item The fractional energy dissipation is small, i.e., $\delta \ll 1$.
\end{enumerate}
Because these assumptions are relatively strong, the direct application of our theory to general nonthermal systems is not obvious and further investigations are necessary.

The spatial homogeneousity (\ref{sup:enum:homo}) may be often difficult to be satisfied, since in many nonthermal systems, the heating and dissipating sources are spatially separated from each other.
For example, if the heating source is an injection of the energetic particle to a system, the entrance point of the energetic particles may be localized at the boundary of the system.
In the low-pressure discharges, such as the one we described in the main text, the dominant dissipation process is the collision with walls, which are spatially located at the system boundary.
However, even in such inhomogeneous systems if the mean free path of the particles are comparable with or longer than the system size, the system can be approximated as homogeneous.
This is the case for our low temperature discharge.

The cross sections of many collision processes have the power dependence on collision energy (\ref{sup:enum:rate}), but often in the limited energy range.
For example, the hydrogen atom-atom collision behaves as $\propto E^{-1/3}$ at $E \lesssim 3$ eV, as can be seen in \fref{sup:fig:crosssection}~(b).
On the other hand, the cross section at $E \gg 3$ eV has a different energy dependence.
Therefore, the G-ML distribution should be only observed at $E \lesssim 3$ eV for hydrogen atoms, although the generation of higher energy atoms has been discussed and observed in a variety of plasmas~\cite{petrovic_excitation_1992,Cvetanovic2009}.  

If the heavy-particle energy is much smaller than the typical energy $\epsilon_0$, 
the zero-temperature assumption (\ref{sup:enum:zero}) may be satisfied.
Even if the heavy-particle temperature is finite, this effect may be corrected by convoluting the Maxwell distribution to the G-ML distribution, as we did in the main text.

\section{Numerical evaluation of G-ML distribution}

Although closed analytical forms are not available for the G-ML distribution, a convenient mixture representation has been reported~\cite{haubold_mittag-leffler_2011,barabesi_new_2016,korolev_mixture_2020},
\begin{align}
    f_{\operatorname{G-ML}}(E) = 
    \frac{1}{\pi} \frac{\nu^{1/\alpha}}{\epsilon_0}
    \int_0^\infty \frac{
        \exp\left(-y{\nu}^{1/\alpha}\frac{E}{\epsilon_0}\right)
        \sin\left(\pi\alpha\nu F_\alpha(y)\right)
    }{
       \left(y^{2\alpha} + 2 y^\alpha \cos(\pi \alpha) + 1\right)^{2\nu}
    }
    dy,
    \label{sup:eq:gml_mixture}
\end{align}
with 
\begin{align}
    F_\alpha(y) = 1 - \frac{1}{\pi\alpha}\cot^{-1}\left(
        \cot(\pi\alpha) + \frac{y^\alpha}{\sin(\pi\gamma)}
    \right).
\end{align}
We evaluate values of the G-ML distribution by numerically integrating \eref{sup:eq:gml_mixture}.

The velocity distribution corresponding to G-ML distribution can be evaluated by substituting $E = \frac{1}{2m}(v_x^2 + v_y^2 + v_z^2)$ and integrate it over $v_x$ and $v_y$ by taking the statistical weight of the space into account.
Since only the term depending on $E$ in \eref{sup:eq:gml_mixture} is $\exp(-cE)$ with $c = y \nu^{1/\alpha}/ \epsilon_0$, we only need the integration of this term.

Let us consider the energy distribution $g(E | c) = c\exp(-cE)$ in 3-dimensional space.
Since the statistical weight of the space is $\sqrt{2mE}$, the distribution of $v_z$ is
\begin{align}
    g(v_z | c)
    &= c\int_{-\infty}^\infty \exp(-cE) \frac{1}{\sqrt{2mE}}dv_x dv_y
    \\
    &= \frac{1}{2}\sqrt{\frac{c}{2m}}\;\Gamma\left(\frac{1}{2}, c\frac{v_z^2}{2m}\right),
    \label{sup:eq:exponential_vdf}
\end{align}
where $\Gamma(s, x) = \int_x^\infty t^{s-1}e^{-t}dt$ is the lower incomplete gamma function.
By substituting \eref{sup:eq:exponential_vdf} into \eref{sup:eq:gml_mixture}, we obtain the velocity distribution of particles with the kinetic energy following the G-ML distribution.

For $x \neq 0$ case, the energy distribution can be obtained by simply multiplying $E^{-x}$ to \eref{sup:eq:gml_mixture}.
\begin{align}
    f(E) \propto 
    \frac{1}{\pi} \frac{\nu^{1/\alpha}}{\epsilon_0}
    \int_0^\infty \frac{
        E^{-x}\exp\left(-y{\nu}^{1/\alpha}\frac{E}{\epsilon_0}\right)
        \sin\left(\pi\alpha\nu F_\alpha(y)\right)
    }{
       \left(y^{2\alpha} + 2 y^\alpha \cos(\pi \alpha) + 1\right)^{2\nu}
    }
    dy.
    \label{sup:eq:gml_mixture_x}
\end{align}
The corresponding velocity distribution can be obtained by replacing $g_x(E | c) = cE^{-x}\exp(-cE)$ by
\begin{align}
    g_x(v_z | c)
    &= \frac{c^x}{2}\sqrt{\frac{c}{2m}}\;\Gamma\left(\frac{1}{2}-x, c\frac{v_z^2}{2m}\right).
    \label{sup:eq:exponential_vdf2_x}
\end{align}

\section{Experimental Details}

\subsection{Spectroscopic Observation of a Commercial Hydrogen Gas Discharge Tube}

\begin{figure*}
    \centering
    \includegraphics[width=15cm]{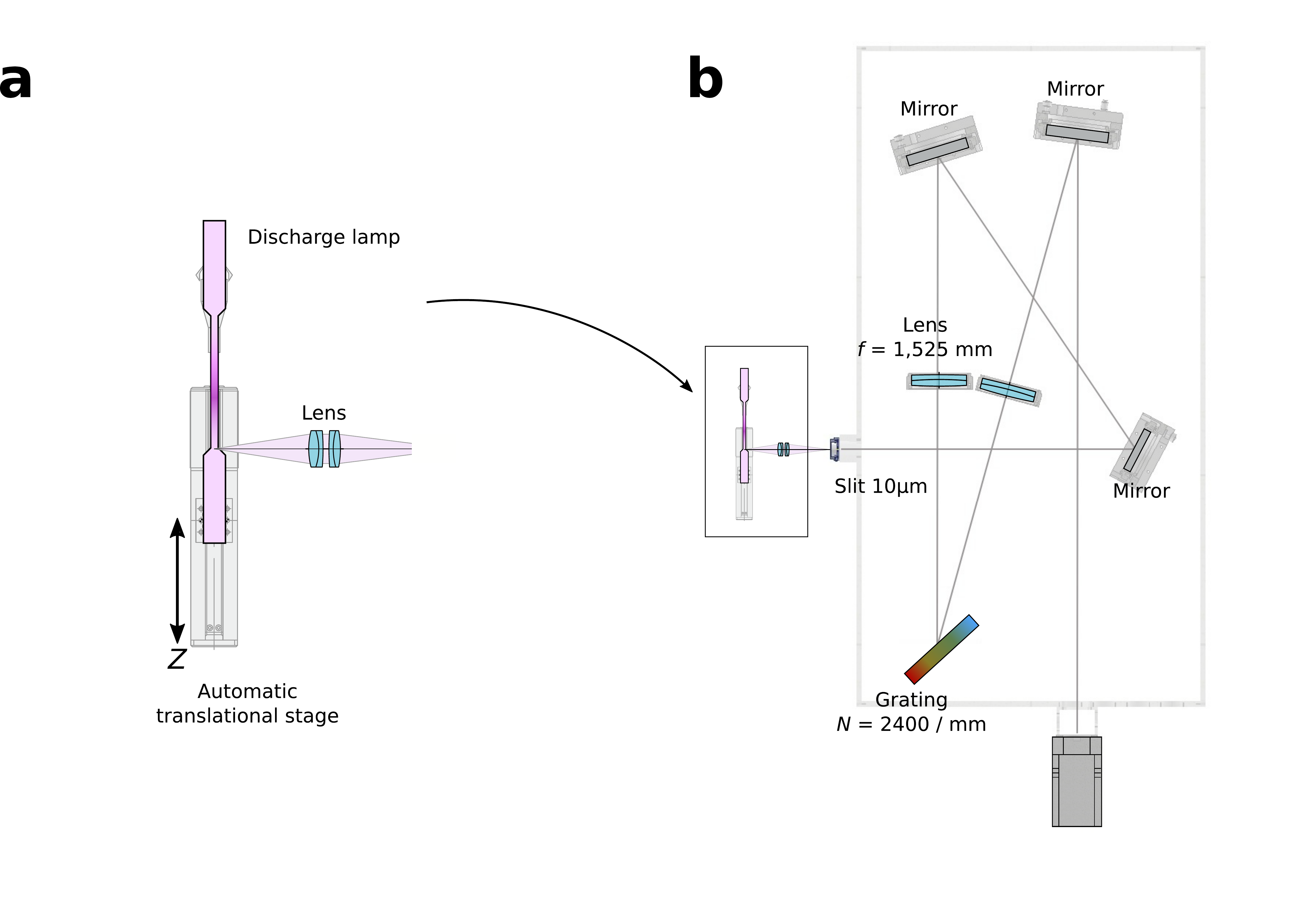}
    \caption{
        A schematic illustration of the experimenta setup for the hydrogen plasma measurement.
        (a) A commercially-available discharge lamp is mounted on an automatic translational stage.
        The emission from the lamp is collected by a pair of achromatic lenses.
        (b) A schematic illustration of the high-resolution spectrometer.
    }
    \label{sup:fig:instrument}
\end{figure*}

\Fref{sup:fig:instrument} shows a schematic illustration of our experimental system to observe hydrogen velocity distribution.
A commercially available hydrogen discharge lamp (Edmund optics, \#60-906) is connected to a DC high-voltage power supply (Spellman SP16122A).
The discharge lamp is mounted on an automated translational stage (SUS Corporation, XA-28L-50E).
The light from the discharge lamp is collected by a pair of achromatic lenses (focal lengths of 80 mm and 100 mm) and focused on the entrance slit of a homemade spectrometer.

In this spectrometer, the light entering through the slit is collimated by an achromatic lens (1525 mm focal length, Edmund optics \#54-568) and dispersed by a diffraction grating (2400 grooves/mm, Jovin Ybon).
The diffracted light is focused by another achromatic lens (1525 mm focal length, Edmund optics \#54-568) on the image sensor of a cooled CCD camera (Andor DV435).
The diffraction grating is mounted on an automated rotational stage (OSMS-60YAW, Sigma Koki)
so that the observed wavelength can be chosen remotely.
Also, the pair of achromatic lenses are mounted on the automated translational stage (HPS60-20x, Sigma Koki) to finely adjust the position and compensate the residual wavelength dependence of their focal lengths.
The resolution was estimated by an iron emission line as 8 pm at the wavelength $\lambda \sim 650$ nm.

\begin{figure*}
    \centering
    \includegraphics[width=6cm]{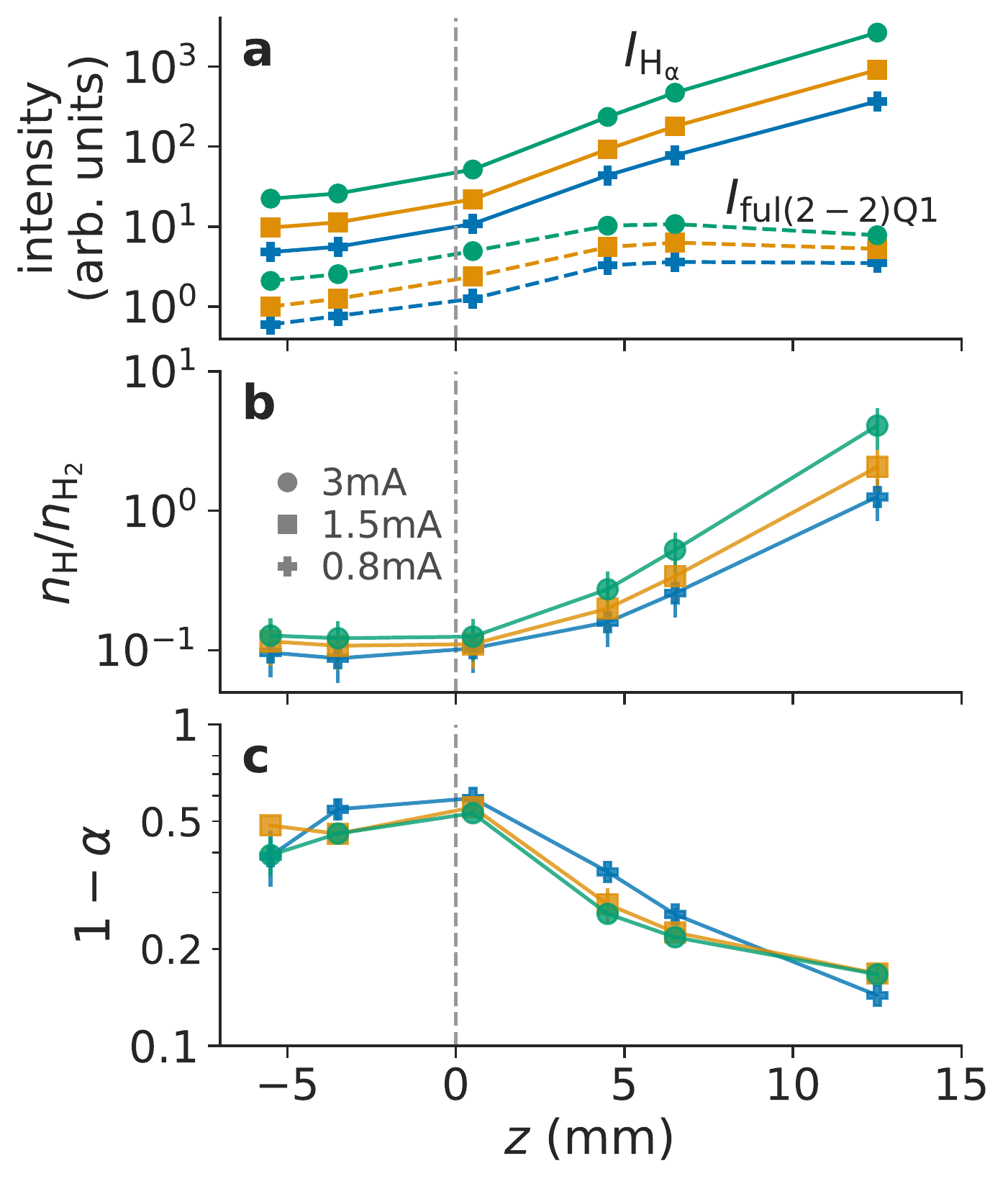}
    \caption{
        Observation result of the hydrogen low-pressure discharge tube.
        (a) The observed intensity of hydrogen atomic Balmer-$\alpha$ line (solid lines) and molecular Fulcher-$\alpha$ $Q1$ line of the $v'-v''=2-2$ transition (dotted lines). 
        (b) The estimated density ratio based on Ref.~\cite{lavrov_determination_2006}
    }
    \label{sup:fig:intensity}
\end{figure*}

We observed emission from the discharge lamp with 0.8, 1.5, and 3 mA discharge current and from several different positions.
In \fref{sup:fig:intensity}~(a), we show the observed intensities of hydrogen atomic Balmer-$\alpha$ line and molecular Fulcher-$\alpha$ $Q1$ line of the $v'-v''=2-2$ transition. 
The edge of the capilarly (1 mm diameter) is at $z = 0$ mm, while in $z < 0$ mm the inner diameter of the discharge tube is $\approx 10$ mm.
Both the intensities are almost constant in $z \lesssim 0$ mm, while in $z \gtrsim 0$ mm the Balmer-$\alpha$ becomes stronger while the Fulcher-$\alpha$ line intensity remains roughly the same.

Several groups have reported a convenient method to estimate the dissociation ratio of hydrogen from the emission line intensities~\cite{lavrov_determination_2006,fantz_spectroscopypowerful_2006}.
We use the result by Lavrov et al.~\cite{lavrov_determination_2006}, where the dissociation ratio is estimated from the line intensities of Balmer-$\alpha$ ($I_\mathrm{H_\alpha}$) and Fulcher-$\alpha$ $Q1$ line of the $v'-v''=2-2$ transition ($I_\mathrm{ful(2-2)Q1}$). 
Although the value of the electron temperature $T_\mathrm{e}$ is necessary to accurately estimate the dissociation ratio, which is not available in our discharge, we assume $T_\mathrm{e} =10^0$ -- $10^1$ eV, which may be reasonable for typical glow discharges.
Based on these values, the density ratio can be approximated by $\approx \eta I_\mathrm{H_\alpha} / I_\mathrm{ful(2-2)Q1}$ with the proportional coefficient $0.5 \times 10^{-2} \lesssim \eta \lesssim 2 \times 10^{-2}$.

In \fref{sup:fig:intensity}~(b), we show the dissociation ratio estimated from the line intensity.
The error bar represented in the figure reflects the uncertainty in the value of $\eta$.
The dissociation ratio starts to increase at the capillary edge.
This tendency can be explained by the small diffusion coefficient of atoms and molecules in the capillary.
Since on the glass surface the association rate of atoms to molecules is smaller than that on the metal surface~\cite{wood_extension_1920}, most of the molecules may be generated on the electrodes located outside the capillary.
On the other hand, the atoms are mainly generated through electron-impact dissociation inside the capillary.
Therefore, the dissociation ratio is higher in the deeper inside the capillary.
From the pressure balance, the atom density in the discuarge tube $n_\mathrm{H}$ is estimated by $\propto\eta I_\mathrm{H_\alpha} / (\eta I_\mathrm{H_\alpha} + I_\mathrm{ful(2-2)Q1})$.

In \fref{sup:fig:intensity}~(c), we show the value of $\alpha$ that gives the best fit to the observed Balmer-$\alpha$ profile.
This value is almost constant outside the capillary and decreases deeper inside the capillary.
This is consistent with the behavior of the dissociation ratio, i.e., because of the higher dissociation ratio in the capillary the rate of the self-collisions of atoms become more significant.

\subsection{Velocity Distribution of Ground-State Hydrogen Atoms in Plasmas}

\begin{figure*}
    \centering
    \includegraphics[width=8cm]{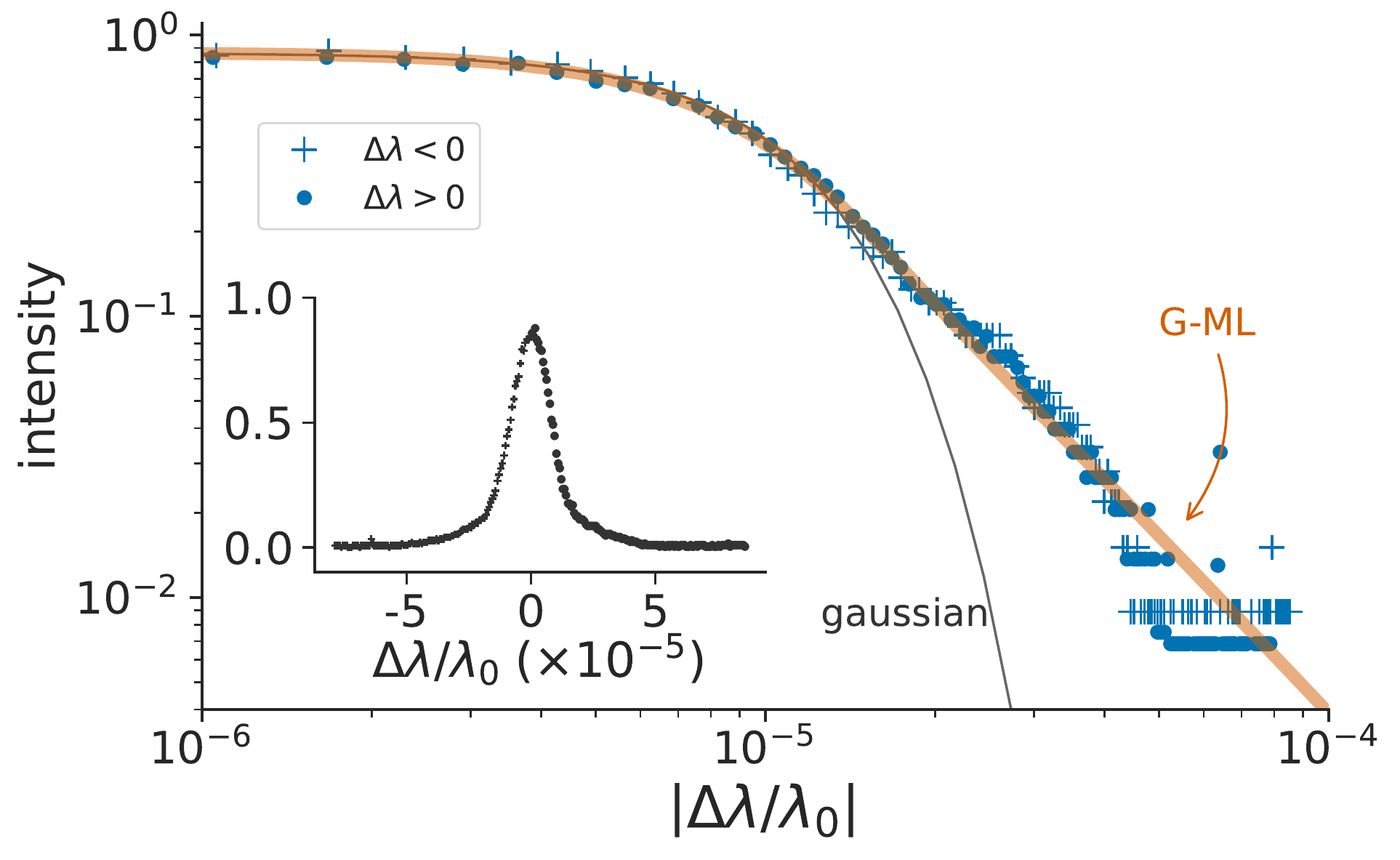}
    \caption{The laser-induced fluorescence spectrum of ground-state hydrogen atoms reported by Amorim et al.~\cite{Amorim1996, Amorim2000}.
    The inset shows the same spectrum in the linear scale.
    The thin and bold curves show the best fit by the Gaussian and Gaussian-convoluted G-ML distribution, respectively.}
    \label{sup:fig:amorim}
\end{figure*}

In the above analysis, we observed the velocity distribution of the excited hydrogen atoms with assuming that it reflects that of the ground state atoms, although the direct generation of excited atoms via dissociative excitation~\cite{McNeill1982,Baravian1987} may make these two distributions different.
In contrast to our experiment, Amorim et al. have directly measured the velocity distribution of ground-state hydrogen atoms in a discharge tube with hydrogen--nitrogen mixture by the two-photon laser fluorescence method~\cite{Amorim2000}.
We extracted the data from their electronic manuscript, where the data are embedded as an \texttt{xml} format.
Their spectrum is shown in \fref{sup:fig:amorim} on a double-logarithmic scale.
The Doppler profile of the ground-state hydrogen atoms also exhibits a power-law dependence in its tails.
From the fit with by the Gaussian-convoluted G-mL distribution (the bold curve in \fref{sup:fig:amorim}), we find $\alpha = 0.42 \pm 0.02$.
From Eq.~(\EqAlpha), a significant dissipation in this plasma compared with the self-collision is suggested.

\subsection{Details of the NSO spectrum}
\begin{figure*}
    \centering
    \includegraphics[width=16cm]{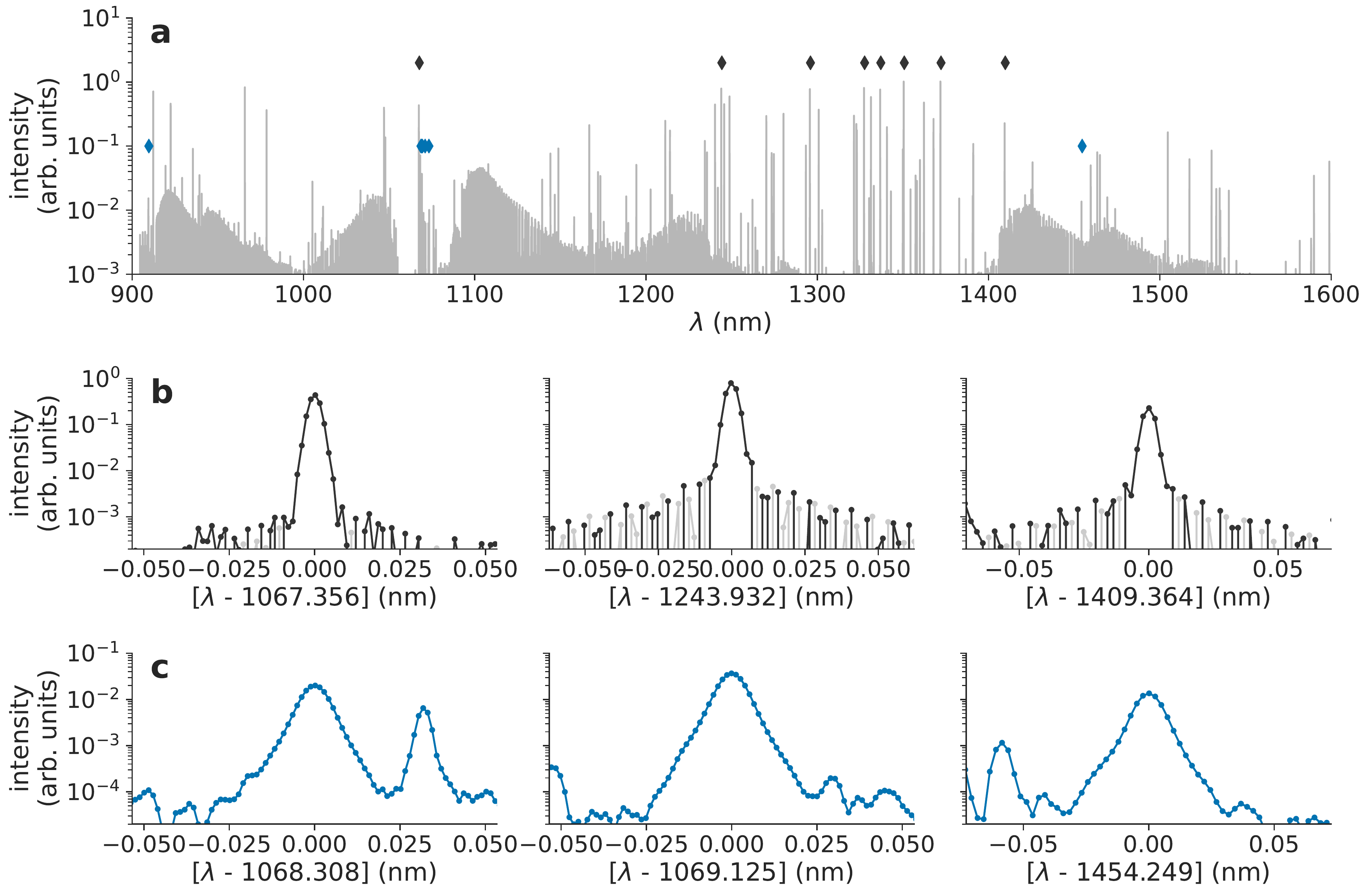}
    \caption{(a) Spectrum observed in NSO~\cite{NSO}. 
    Markers show the wavelengths of the argon and carbon lines analyzed in the main text. 
    (b) Expanded spectra of three argon lines.
    The light gray markers show the negative points (multiplied by -1), representing the oscillation of the instrumental side robes.
    (c) Expanded spectra of three carbon lines.
    }
    \label{sup:fig:nso}
\end{figure*}

In the main text, we analyzed a high-resolution spectrum by National Solar Observatory~\cite{NSO}.
This was originally measured to study cyanide spectra in 1977 from a microwave discharge with the mixture of carbon, nitrogen, and argon.
This measurement was carried out with a high-resolution Fourier transform spectrometer with a 1-m optical path difference in the wavelength range $\lambda=$ 830--2710 nm. 
The wavelength resolution is 3.5 pm (as the full width half maximum) for this measurement.
\Fref{sup:fig:nso}~(a) shows the original data and \fref{sup:fig:nso}~(b) and (c) show expanded views for argon lines at $\lambda =$ 1350, 1372, and 1337 nm
and
carbon lines at $\lambda =$ 1069.4, 1068.6, and 1454 nm, respectively.

The observed line width of the argon lines is 5.2 pm. 
This is consistent with the convolution of the argon Doppler width at the room temperature (3.7 pm) and the instrumental width.
The oscillation seen in the argon line wings, i.e., the side robe, originates from the instrumental function of this spectrometer.
Since this spectrometer is based on the Fourier transform, the spectrum is affected by the window function, such as the sinc function $\frac{\sin\left(\pi \Delta \lambda / \delta_\lambda \right)}{\pi \Delta \lambda / \delta_\lambda}$ for a rectangular window.
Although other line broadenings, such as the Doppler broadening, averages out this oscillation in the instrumental side robes, this is still apparent in the argon lines because of their similar Doppler widths to the instrumental width. 
However, this effect is negligible for the carbon lines owing to their much larger Doppler widths.

\end{document}